\documentclass{article}

\usepackage{arxiv}

\usepackage[utf8]{inputenc} 
\usepackage[T1]{fontenc}    
\usepackage{hyperref}       
\usepackage{url}            
\usepackage{booktabs}       
\usepackage{amsfonts}       
\usepackage{nicefrac}       
\usepackage{microtype}      
\usepackage{lipsum}
\usepackage{graphicx, caption, subcaption}
\usepackage[abbreviations,symbols,nogroupskip,nonumberlist]{glossaries-extra}

\usepackage{tabularx}
\usepackage{algorithm,algorithmicx,algpseudocode}
\usepackage{xcolor}
\usepackage{multirow}
\usepackage{comment}

\newcommand{\myparagraph}[1]{\vspace{10pt}\noindent\textbf{#1}}
\newcommand{\todo}[1]{\textcolor{red}{#1}}
\newcommand{\mobixim}{\emph{MobiXIM}}

\title{A Framework for Devising, Evaluating and Fine-tuning Indoor Tracking Algorithms}

\author{
 Alpha Diallo \\
  Department of Information Systems\\
  University of Lausanne, Switzerland\\
  \texttt{alpha.diallo@unil.ch} \\
   \And
 Benoit Garbinato \\
  Department of Information Systems\\
  University of Lausanne, Switzerland\\
  \texttt{benoit.garbinato@unil.ch} 
}

\begin{document}
\maketitle

\begin{abstract}

In recent years, we have observed a growing interest in Indoor Tracking Systems (ITS) for providing location-based services indoors. This is due to the limitations of Global Navigation and Satellite Systems, which do not operate in non-line-of-sight environments.
Depending on their architecture, ITS can rely on expensive infrastructure, accumulate errors, or be challenging to evaluate in real-life environments.
Building an ITS is a complex process that involves devising, evaluating and fine-tuning tracking algorithms. This process is not yet standard as researchers use different types of equipment, deployment environments, and evaluation metrics.
Therefore, it is challenging for researchers to build novel tracking algorithms and for the research community to reproduce the experiments.

To address these challenges, we propose \emph{MobiXIM}, a framework that provides a set of tools for devising, evaluating and fine-tuning tracking algorithms in a structured manner.
For devising tracking algorithms, \mobixim{} introduces a novel plugin architecture, allowing researchers to collaborate and extend existing algorithms. 
We assess our framework by building an ITS encompassing the key elements of wireless, inertial, and collaborative ITS. 
The proposed ITS achieves a positioning accuracy of 4 m, which is an improvement of up to 33\% compared to a baseline Pedestrian Dead Reckoning algorithm.

\end{abstract}

\section{Introduction}

\emph{Indoor Tracking} refers to solutions that overcome the limitations of Global Navigation and Satellite Systems (GNSS), such as the Global Positioning System (GPS) and Galileo, in non-line-of-sight environments.
It is becoming a trendy topic in the industry and the research community because of its potential impact on people's lives.
To track mobile devices indoors, researchers propose solutions built with technologies whose primary goals were not specifically designed for indoor tracking. These solutions can be classified into three categories described hereafter.

\myparagraph{Infrastructure-based ITS}, defined as \emph{wireless-based}, rely on existing or dedicated communication infrastructure mainly using a centralized architecture by offloading computationally extensive tasks to a remote server.
In this approach, the tracking is done by measuring a Received Signal Strength Indication (RSSI) between an emitter and a receiver to either estimate distance or capture signal fingerprints. 
RSSI is sensitive to signal variation caused by multipath, fading, reflection or signal scattering. 
To improve the accuracy of wireless-based ITS, researchers use multiple techniques such as trilateration~\cite{bembenik_ble_2020}, multilateration~\cite{adhikari_neural_2021}, map matching~\cite{zhang_adaptable_2020} or fingerprinting~\cite{torres-sospedra_scalable_2022}.
These techniques leverage signals from nearby devices or use landmarks to track mobile devices~\cite{jang_survey_2023}. 
For instance, fingerprinting, popular across the literature, requires researchers to collect signals at predefined locations and compare them to signals received by a mobile device to estimate its location. 
This approach is challenging to implement as it requires regular data collection and a high cost for deploying and maintaining the infrastructure.

\myparagraph{Infrastructure-less ITS}, defined as \emph{inertial—and magnetic-based} ITS, rely on inertial and magnetic sensors embedded in recent mobile devices. These sensors measure physical activity or the magnetic field.
The accelerometer measures a linear acceleration, usually on three axes, the gyroscope measures an angular velocity, and the magnetometer measures the strength and direction of the Earth's magnetic field~\cite{kunze_can_2010}.
This approach is commonly used in decentralized ITS as mobile devices can compute their location without relying on a central server~\cite{diallo_decentralized_2024}. 
However, these ITS suffer from an accumulation of errors due to noisy sensors affecting their tracking estimates.

\myparagraph{Collaborative ITS} is a novel approach to overcome the limitations of the above-listed type of ITS. It consists of leveraging the communication capability of mobile devices to sense their environment and exchange information with nearby mobile devices~\cite{diallo_decentralized_2024}. 
In some configurations, this approach can be combined with a limited number of fixed infrastructures to improve accuracy.

\subsection{Scope and Methodology}
This paper focuses on the three types of ITS, specifically those deployed on \emph{digital consumer electronics} such as smartphones, tablets, microcontroller units (MCUs), etc.
Therefore, we exclude industrial or military tracking systems requiring expensive infrastructure.

ITS within our scope of interest are built using principles centred around devising, evaluating, and fine-tuning the tracking algorithms.
Given the lack of standards for devising tracking algorithms, it is difficult for researchers to collaborate and reuse existing algorithms.
Additionally, most of these algorithms are evaluated on data collected by the researchers. 
Collecting data is a time-consuming task that requires multiple devices, well-planned coordination between participants, and an infrastructure for storing and analysing the data.

We also observe that ITS are evaluated using different metrics, making comparisons between them challenging. Additionally, the lack of a framework hinders the reuse of existing tracking algorithms and complicates the reproduction of experiments.
Regarding reproducibility, a survey shows that more than 70\% of researchers have tried and failed to reproduce another scientist's experiments, and even more than half have failed to reproduce their own experiments~\cite{baker_1500_2016}.
Reproducibility is a major concern in ITS as experiments are conducted in specific environments, use costly infrastructure, and require timely interactions between mobile devices and meticulous coordination between participants. 
Indeed, once the experiments are done, it is difficult for other researchers to reproduce because of the absence of data and the difficulty of replicating the environment in which the experiments were achieved.

\subsection{Problem statement}
This paper addresses the problem of standardizing the process of devising, evaluating, and fine-tuning tracking algorithms. Setting up a clear methodology and providing tools are key to accelerating the process of building ITS and facilitating the reproducibility of experiments.

To achieve these objectives, we propose a novel framework that defines a standard set of processes for collecting data, devising, evaluating and fine-tuning tracking algorithms.
Our framework integrates an orchestrator platform for preparing the data collection, processing the data and replaying the movement of participants. A mobile companion app is also proposed, allowing participants to follow instructions to collect data along selected paths defined using the orchestrator platform.
At its core, \mobixim{} uses an extensible plugin architecture, allowing researchers to devise specific parts of their tracking algorithm and implement them on top of existing state-of-the-art algorithms. This methodology, commonly used in software engineering, accelerates the development of prototypes by removing the complexity of non-core components of a system.

To the best of our knowledge, we are the first to propose a complete framework for devising and evaluating the three types of ITS.

\subsection{Roadmap}
The remainder of this paper is organized as follows.
In the next section, we present the related work and explain the need for a framework encompassing most ITS.
In Section~\ref{sec:model}, we place the context and define some key terms.
We present the methodology used in the literature in Section~\ref{sec:methodology}.
Section \ref{sec:framework} introduces \mobixim{} and describes each component.
In Section~\ref{sec:experiments}, we run experiments by building an ITS. We evaluate this ITS built using \mobixim{} in Section~\ref{sec:evaluation}.
In Section~\ref{sec:discussion}, we discuss the strengths and limitations of \mobixim{} and conclude the paper in Section~\ref{sec:conclusion}. 
\section{Related work}
\label{sec:related}


This section presents recent frameworks for devising indoor tracking algorithms. We introduce these frameworks, discuss their main contributions and show how they compare to \mobixim{}. We summarise our findings in Table~\ref{tab:related}.

\begin{table}[ht]
    \caption{Comparing the features of MobiXIM and other frameworks}
    \vspace{7pt}
    
    \centering
    \begin{tabular}{|p{2.4cm}|p{2.3cm}|p{2.3cm}|p{2.3cm}|p{2.3cm}|}
    \hline
         & MobiXIM & Ko and Wu~\cite{ko_framework_2022} & Chen et al.\cite{chen_novel_2021} & De Wynckel and Signer~\cite{de_wynckel_indoor_2021}\\\hline
        Wireless-based & Yes & Yes& Yes & Yes \\ \hline
        Inertial-based & Yes & No & No & Yes \\ \hline
        Collaborative & Yes & No & No & Yes \\ \hline
        Execution replay & Yes & No & No & No \\ \hline
        Evaluation metrics & Trajectories similarity and Positioning Accuracy & Classification accuracy and Positioning error (MSE) & Positioning Accuracy & Positioning Accuracy\\ \hline
        Type of data & Simulated and Real-life & Simulated and Real-life & Real-life & Real-life\\ \hline
        Wireless Protocol & BLE & Wi-Fi & BLE & Wi-Fi, BLE, RFID, LTE\\ \hline
        Multi-user Platform & Yes & No & No & No \\ \hline
        Type of trackees & Smartphones and tablets & Laptops & Microcontrollers & Smartphone, tablets, MCUs, laptops, etc. \\ \hline
        Floorplan representation & GeoJSON & Local coordinate frame with image overlay & Local coordinate frame with image overlay & Not specified \\ \hline
    
    \end{tabular}
    
    \label{tab:related}
\end{table}

In the literature, most researchers focus on wireless-based ITS and propose frameworks for improving the processes for devising ITS based on fingerprinting~\cite{brunello_framework_2022,song_novel_2019,ko_framework_2022}.
One such framework, proposed by Ko and Wu, incorporates channel modelling, position estimation, and error analysis methods for wireless-based ITS using RSSI collected from Wi-Fi Access Points (AP)~\cite{ko_framework_2022}. 
Their framework uses two positioning methods to achieve coarse positioning and fine positioning. For coarse positioning, RSSI are partitioned into clusters by their source spaces. Then, they use a Support Vector Machine (SVM) to classify the RSSI and find the corresponding room where a mobile device is located. 
They use a Bayesian estimation technique for fine positioning to pinpoint a mobile device's location based on its previously estimated coarse location.
They achieve a 99.1\% accuracy for estimating a coarse location and about 3 meters of positioning accuracy for fine positioning in a 608 m\textsuperscript{2} environment and a density of 2.1 Wi-Fi AP every 100 m\textsuperscript{2}. 
Their framework's main limitation is its limited scope, focusing solely on devising tracking algorithms using fingerprinting.
ITS based on fingerprinting are challenging to deploy in real-life environments because they require time-consuming data collection, and their performance depends strongly on environmental changes. In addition, the framework proposed by Ko and Wu uses a classification model based on SVM coupled with Bayesian estimation methods, which requires significant computational resources. Therefore, their proposed techniques are unsuitable for real-time applications. 

Chen et al. propose a framework for devising ITS using Machine Learning (ML) algorithms~\cite{chen_novel_2021}. 
They collect Bluetooth signals at predefined locations and pass them to neural networks to estimate the location of mobile devices. Then, they compare the performance of a Multilayer Perceptron (MLP) with a Recurrent Neural Network (RNN) on a dense dataset of collected signals. They demonstrate that the MLP outperforms the RNN and offers an accuracy of up to 98\% with six receivers (a density of 8.3 receivers per 100m\textsuperscript{2}). 
Their proposed framework addresses the challenges of finding an optimal algorithm to solve the environmental factors affecting indoor positioning.
However, as this approach requires a lot of data, it is important to consider the challenges of collecting data in a large environment. Furthermore, their framework does not include aspects related to the representation of deployment environments and does not provide guidelines for reproducing the experiments.
The latter limitation is common in the literature, as most authors do not provide enough information to reproduce their experiments. Even if they detail their methodology for devising their algorithms, researchers expect a thorough discussion, code, and datasets to facilitate the reproducibility of the experiments.

De Wynckel and Signer propose OpenHPS, an open-source hybrid positioning system using a modular framework that supports multiple technologies and positioning methods~\cite{de_wynckel_indoor_2021}. It is designed to be flexible by fusing data from multiple sources, thus integrating wireless and inertial measurements. The collected data can be stored in databases or locally on a mobile device. Their proposed framework is built in TypeScript, a cross-platform superset of JavaScript, ensuring deployment on mobile, client, and server-side applications. 
Their framework is primarily aimed at a community of developers, enabling them to design a hybrid, multi-platform system encompassing indoor and outdoor positioning.
Therefore, they do not address the challenges of evaluating ITS and reproducing experiments.

Other frameworks proposed in the literature come with powerful tools and structured processes to orient and speed up research. An example of a framework for a specific use case is proposed by Kitras et al.~\cite{kitras_location_2023}. They focus on integrating location modules into air quality measurement systems. By proposing a Location and Movement Detection of the Application layer (LaMDA) framework, they challenge researchers to check the reliability of location information provided by low-cost devices for analysing Air Quality.

In the literature, researchers have not yet provided a complete framework that considers all the significant aspects of an ITS, from devising the tracking algorithms to evaluation, emphasising reproducibility. Our proposed framework addresses this, providing guidelines and tools for devising tracking algorithms in a structured manner. 
\section{System model}
\label{sec:model}

We consider environments where an ITS is needed such as undergrounds, buildings with multiple rooms, facilities, corridors, etc.
We aim to track mobile devices with embedded computational and communication capabilities. These mobile devices, typically smartphones and tablets, can sense their environment, collect inertial and magnetic measurements and wireless signals, run some tracking algorithms, and communicate with nearby devices wirelessly.

We also consider Bluetooth Low-Energy (BLE) beacons with fixed positions to correct mobile device estimates.
In this setting, mobile devices detect nearby beacons and estimate their location according to the beacon with the strongest signal or compare the signal fingerprints with priorly collected signals.
Beacons can be physical or virtual. Physical beacons are commercial devices running on batteries and broadcasting advertisements following the Eddystone or iBeacon standards. 
Virtual beacons are simulated devices used to facilitate the execution of scenarios under several configurations without installing and maintaining physical beacons. They emit signals propagating using a path loss model that models the relationship between RSSI and distance. 

The goal of an ITS is to estimate single locations or trajectories by using measurements from mobile devices.  
A trajectory $T$ consists of a set of $n$~tuples $L_{i \in \{1..n\}} = (\lambda_i, \phi_i, t_i)$, where $\lambda_i$ is a latitude, $\phi_i$ is a longitude and $t_i$ is a timestamp.
In this paper, we distinguish three types of trajectories:
\begin{itemize}
    \item \textbf{Groundtruth trajectory} is a sequence of points representing the real locations of users at a given time.
    \item \textbf{Estimated trajectory} represents locations as computed by a tracking algorithm associated with a baseline tracking algorithm.
    \item \textbf{Corrected trajectory} is the trajectory resulting from improving the estimated trajectory.
\end{itemize}

To obtain the estimated and corrected trajectories, researchers devise tracking algorithms that can fit into one of the following categories.
\begin{itemize}
    \item \textbf{Filtering algorithm} is used to remove the noise from collected data or to smooth a signal. One such algorithm is a low-pass filter used to smooth rapid fluctuation of RSSI in wireless-based ITS or to smooth inertial measurements that go beyond a given threshold~\cite{mehrabian_sensor_2023,diallo_decentralized_2024}.
    \item \textbf{Positioning algorithm} estimates the location of a device using the data processed by the filtering algorithm. For inertial-based ITS, the baseline positioning algorithm is the Pedestrian Dead Reckoning (PDR), which estimates a new location based on the previous location coupled with the orientation of a user and its step length. 
    For wireless-based ITS, k-nearest neighbors (k-NN) is a well-known technique for detecting the closest priorly collected signal to a newly detected RSSI in a fingerprinting approach.
    \item \textbf{Collaborative algorithm} is used to further improve the location estimates of a positioning algorithm by leveraging the proximity between users. 
\end{itemize}

\section{Methodology}
\label{sec:methodology}

This section presents the common steps researchers take to build new ITS. These steps are part of a methodology that \mobixim{} aims to simplify and standardise.

\subsection{Devising tracking algorithms}
\label{subsec:methodology-devising}
Most tracking algorithms proposed in the literature are enhancements of existing algorithms, such as the PDR commonly used for inertial-based ITS or k-NN for finding nearest neighbours in wireless-based ITS. These algorithms are popular and have been implemented multiple times. Rewriting them requires a lot of time and can even introduce errors due to a wrong implementation, thus impacting the performance of the tracking algorithms. 
Researchers must reuse existing algorithms and assemble them easily to reduce the hassle of devising new tracking algorithms.

\subsection{Evaluating tracking algorithms}
\label{subsec:methodology-evaluating}
After devising and implementing tracking algorithms, researchers must evaluate their performance and compare them with existing algorithms. 
In the literature, ITS are evaluated regarding positioning accuracy, coverage, complexity, robustness, scalability, cost, privacy and power consumption~\cite{mendoza-silva_meta-review_2019}. 
However, positioning accuracy remains by far the most used evaluation criterion. Rainer Mautz defines positioning accuracy as the degree of conformance of an estimated or measured position at a given time to the true value~\cite{mautz_indoor_2012}. 
Therefore, researchers assess the performance of the tracking algorithms by comparing the corrected trajectories with the corresponding groundtruth and estimated trajectories.
They use multiple metrics, such as the Mean Squared Error (MSE), the Mean Absolute Error (MAE), or the Root Mean Squared Error (RMSE).
However, these metrics are sensitive to outliers and do not measure the similarity between the groundtruth and the corresponding estimated and corrected trajectories.
It is important to define evaluation metrics that would be used as standards in the literature to compare ITS better. 
After selecting the metrics for measuring the performance of their tracking algorithms, researchers evaluate them using one of the approaches described hereafter.

\subsubsection{Evaluating tracking algorithms on existing data.} 

Some researchers use existing or synthetic datasets to evaluate their tracking algorithms.
Existing data are public mobility datasets or datasets initially collected in previous experiments. 
Synthetic datasets are generated during a simulation to mimic the real movements of users in an indoor environment~\cite{huang_indoorstg_2013,li_vita_2016}.
Such datasets accelerate the evaluation of the proposed tracking algorithms. 
However, in real-life environments, the performance of the tracking algorithms may diverge as environments have different layouts, which affect the raw measurements and the resulting estimated and corrected trajectories.
On the other hand, synthetic datasets can be biased or unrealistic, thus failing to capture the real movements of people in indoor environments~\cite{diallo_mobixim_2023}.

\subsubsection{Evaluating tracking algorithms on collected data.}
Another method for evaluating an ITS involves collecting data. While this approach provides greater flexibility for the researcher, it requires completing the following steps before evaluating the tracking algorithms.

\begin{itemize}
    \item \myparagraph{Planning the data collection.}
    Before collecting the data, researchers need to understand their targeted deployment environment. This involves knowing the dimensions of the environment, its occupancy rate, constraints related to the architecture of the building, and the layout of the furniture. 
    %
    These parameters can impact the technology choice and the performance of the tracking algorithms.

    \item \myparagraph{Building a data collection app.}
    In the literature, most authors use ad-hoc software tools to collect data.
    For instance, Jimenez et al. propose \emph{GetSensorData}, an Android app for collecting data from wireless, inertial and magnetic sensors~\cite{jimenez_tools_2019}. However, the code source is no longer maintained to consider the updates from the Android Operating System. 
    Using a standard mobile application to collect raw measurements will reduce the time and effort needed to evaluate ITS. To fully benefit from this mobile application, it must be fully integrated into an ecosystem to better coordinate the data collection.
    
    \item \myparagraph{Collecting the data.}
    Data collection is a tedious task usually done by a small group of participants. 
    In the literature, most ITS are evaluated in small deployment environments, generally less than 500 m\textsuperscript{2}, and with a small number of trajectories.
    %
    This is mainly due to the difficulties of coordinating teams for large-scale data collection. 
    For example, in collaborative ITS, researchers must capture participants' interactions during the data collection. Since collaborative ITS perform better with a high number of interactions between participants, the data collection process must involve multiple participants moving simultaneously~\cite{diallo_decentralized_2024}.

\end{itemize}

Once the evaluation metrics are set and the data are ready, researchers proceed with the optional steps outlined below. 

\subsubsection{Designing the floor plans.}
After selecting an indoor environment, the next step is to model it to facilitate the data collection and visually display the trajectories.
In the literature, most researchers use georeferenced images to model their indoor environment\cite{ko_framework_2022,chen_novel_2021,werner_efciently_2011}.
This approach is tedious to implement and difficult to update when the building changes. It also requires integrating properties of the environment, such as rooms, walls, or doors. 

To facilitate the widespread adoption of an ITS, researchers should integrate dynamic maps that consider the specific features of the environment. 
Additionally, the map should use a geographic coordinate system to facilitate its integration into existing outdoor positioning systems.

\subsubsection{Cleaning the data.}
Raw measurements collected by digital consumer electronics are noisy and can contain outliers impacting ITS performance.
Some authors use filtering algorithms to smooth the raw measurement, attenuating the noise or removing outliers by eliminating anomalous signals~\cite{ye_low-cost_2019}.
%
Multiple tools exist for data cleaning. One such tool is the SciPy library, available in Python for processing signals.\footnote{\url{https://scipy.org/}}
Instead of devising the filtering algorithms from the ground, researchers can use existing libraries and plug them into their positioning algorithms.

\subsubsection{Adjusting parameters.}
After cleaning the data, researchers may need to adjust the parameters associated with each participant or mobile device. These parameters include the sampling rate, the step length, the initial orientation, the transmission power, and the error correction threshold. 
%
%
It would be helpful to have tools to adjust these parameters and observe how they impact the performance of the tracking algorithms.

The methodology presented in this section demonstrates the numerous challenges researchers face when devising and evaluating their tracking algorithms. The complexity of the processes, coupled with the absence of standards, makes it even more challenging to compare ITS and reproduce the experiments. 
Therefore, it is crucial to have a framework that guides researchers and offers functionalities such as pre-loaded datasets to accelerate devising and evaluating new tracking algorithms. Therefore, such a framework would provide a common baseline for comparing tracking algorithms. 
\section{\mobixim{} Architecture}
\label{sec:framework}

This section describes the components and the process flow of \mobixim{}, associated with the characteristics of ITS listed in Section~\ref{sec:methodology}.
Figure~\ref{fig:architecture} highlights these components and indicates how they interact. %

\subsection{Mobile Companion App}

As discussed in Section~\ref{sec:methodology}, some researchers evaluate their tracking algorithms with collected data that best fits their needs. This task is time-consuming and does not follow any standard protocol. 
\mobixim{} facilitates the data collection with a mobile companion app that integrates seamlessly with other framework components.
Built for iOS and Android, it is intended to be used by participants to collect raw measurements along predefined groundtruths. 
These raw measurements are made of the following fields stored in CSV files.

\begin{itemize}
    \item \textbf{AccX}. It measures the acceleration on the X-axis, corresponding to the left and right horizontal movements.
    \item \textbf{AccY}. It measures the acceleration on the Y-axis, corresponding to the horizontal forward and backward movements.
    \item \textbf{AccZ}. It measures the acceleration on the Z-axis corresponding to the vertical movements up and down.
    \item \textbf{Gyroscope}. It measures the orientation relative to the body frame.
    \item \textbf{Azimuth}. It measures the rotation angle between the device's Y-axis and the magnetic north pole.
    \item \textbf{Pitch}. It measures the rotation angle on the X-axis, the angle between a plane parallel to the device's screen and a plane parallel to the ground.
    \item \textbf{Roll}. It measures the tilt of the device on the Y-axis.
    \item \textbf{RSSI}. It measures the signal strength in decibel-milliwatts (dBm) emitted by physical BLE beacons at a short distance, estimated using the inverse relationship between distance and RSSI. A strong RSSI indicates a short distance between a mobile device and a beacon. A weak RSSI indicates a larger distance. We set a default value of -100 dBm to indicate a beacon that is not within the detection range of a mobile device.
\end{itemize}

\subsection{Orchestrator platform}

\begin{figure}
    \centering
    \includegraphics[width=1\linewidth]{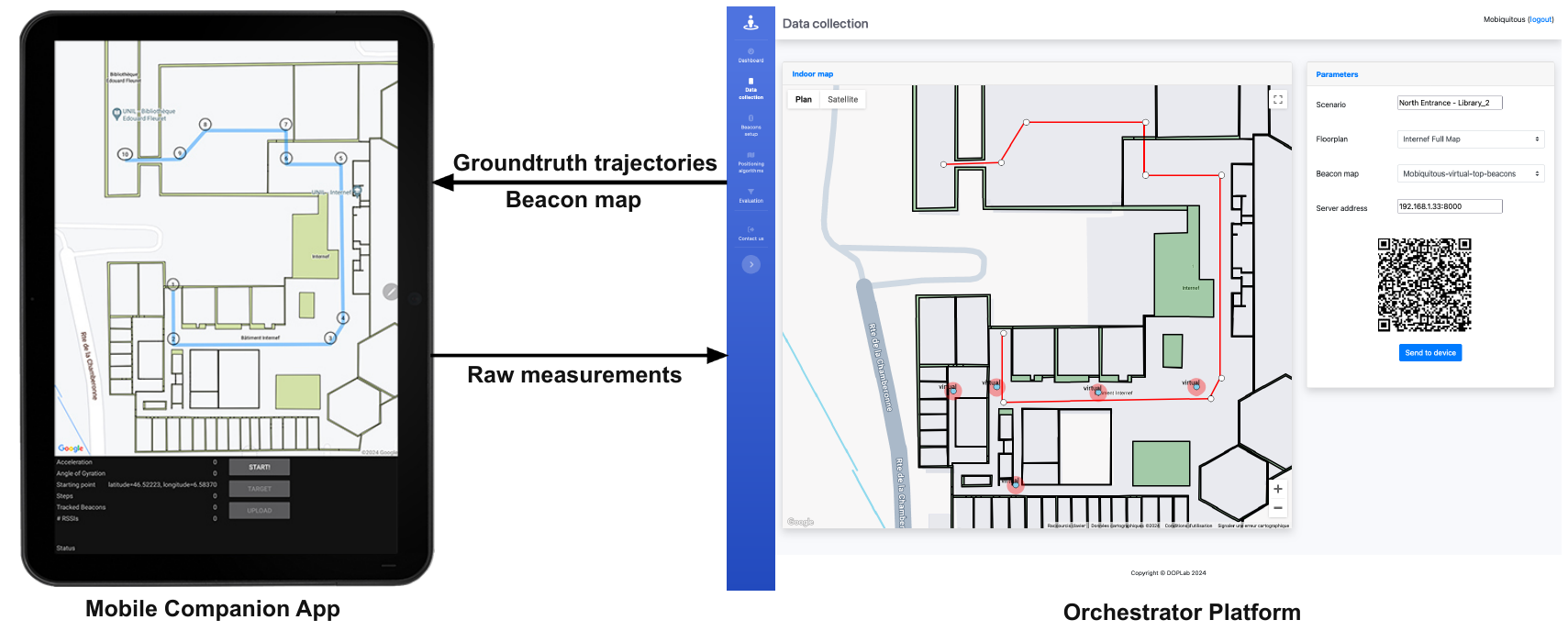}
    \caption{User interfaces of the Mobile Companion App and the Orchestrator Platform}
    \label{fig:companion-platform}
\end{figure}

The orchestrator platform is a web application built in Python using the Django Framework and hosted on a remote server. It is intended to prepare the data collection, set up the experiments, and run the evaluation.
The user interface, mainly built in HTML/CSS and JavaScript, is intended for use on a desktop browser.
Figure~\ref{fig:companion-platform} shows the user interfaces of the mobile companion app and the orchestrator platform. The role of the orchestrator platform and its interaction with the mobile companion app are detailed hereafter.

\begin{figure}[t]
    \centering
    \includegraphics[width=1\linewidth]{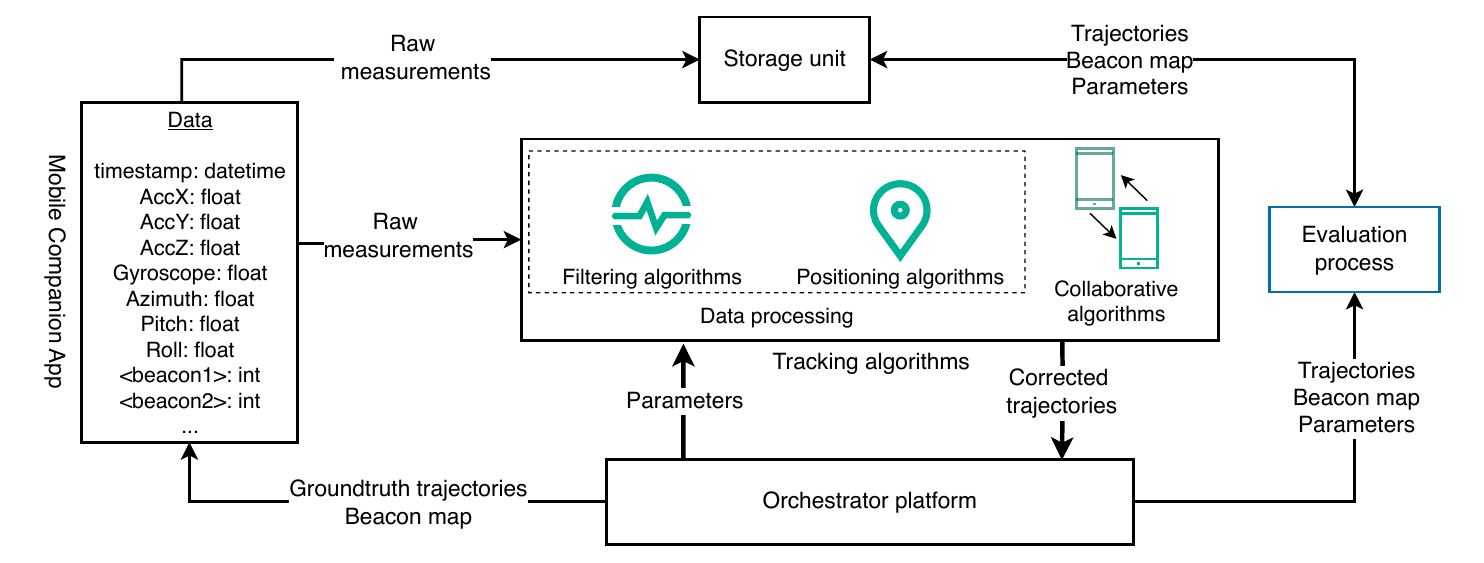}
    \caption{Architecture and process flow of \mobixim{}}
    \label{fig:architecture}
\end{figure}

\myparagraph{Planning scenarios.} Prior to the data collection, as shown in Figure~\ref{fig:architecture}, researchers plan the execution scenario by designing the groundtruth trajectories on the orchestrator platform. These trajectories are then sent to the mobile companion app via a QR code scanned by participants to receive the groundtruths along which they collect data. During the data collection, participants regularly signal when they reach checkpoints using the Mobile Companion App.

\myparagraph{Floorplan representation.} To facilitate the construction of floorplans, we use the GeoJSON format for encoding geographic features.
This format is particularly useful for running spatial queries such as detecting rooms, corridors or intersecting walls. 
In the literature, most authors represent their deployment environment with an image layer in a 2D cartesian representation system~\cite{pascacio_collaborative_2022,mansour_power_2023}.
However, this representation is static and difficult to scale and manipulate.
Representing an indoor space with a GeoJSON format facilitates interoperability with existing outdoor tracking systems due to its scalability.

\myparagraph{Execution Replay.} Another function of the orchestrator platform is to aggregate the raw measurements and groundtruth trajectories into a single environment to reproduce the movement of participants simultaneously. 
We define this approach as \emph{execution replay}, allowing researchers to replay the real displacement of participants.
%
This approach helps researchers to visualize the performance of their tracking algorithms better.
It also allows them to create environments that foster collaboration using real-life data collected by participants.
This approach addresses the challenges of synchronizing participants in real environments where many people exchange data simultaneously. 
Additionally, it significantly reduces the time and cost of running experiments while relying on real-life data reflecting the real movements of people indoors.


\subsection{Tracking algorithms}

The role of a tracking algorithm is to process raw measurements and estimate locations. 
As \mobixim{} aims to facilitate reusability, we structure the tracking algorithms with a plugin architecture.
A plugin architecture is a design pattern used in software engineering to build modular components that are independent of each other.
These components can be placed together without altering the core codebase. 
This novel approach for indoor tracking allows us to separate the main steps for devising an ITS, such as cleaning the data, computing the location estimates, and simulating data exchanges. 
With this approach, researchers can reuse existing filtering and positioning algorithms and build their tracking algorithms on top of them.

In the process flow illustrated in Figure~\ref{fig:architecture}, the raw measurements are first sent to the filtering algorithms for preprocessing. Then, positioning algorithms compute the estimated trajectories and send these trajectories to the collaborative algorithm during an execution replay.


To implement the plugin architecture, we specify software interfaces for each type of algorithm. These interfaces are skeletons that researchers should follow to build their own tracking algorithms. For instance, all the tracking algorithms should implement the basic functions defined below.
\begin{itemize}
    \item \emph{get\_plugin\_name}: returns the full name of a plugin.
    \item \emph{get\_plugin\_slug}: returns a slug, an abbreviated form of the full name used as a unique plugin identifier.
    \item \emph{get\_plugin\_display\_name}: returns the name displayed on the orchestrator platform.
    \item \emph{get\_plugin\_category}: returns one of the three types of algorithms implemented by the plugin, namely, filtering, positioning and collaborative.
\end{itemize}

In addition to these functions, each algorithm possesses a predefined function for data processing. For filtering algorithms, this function is called \emph{get\_filtered\_data} and takes as inputs raw measurements and returns smoothed data.
For positioning algorithms, the function is called \emph{get\_positioning\_data}. It takes as inputs the groundtruths, filtered or raw measurements, the initial location of the device and optional parameters such as the estimated step length, and it returns an estimated trajectory.
Collaborative algorithms are executed only when at least two devices are within a detection range set by the researcher. 
%
%
\begin{figure}
    \centering
    \includegraphics[width=0.5\linewidth]{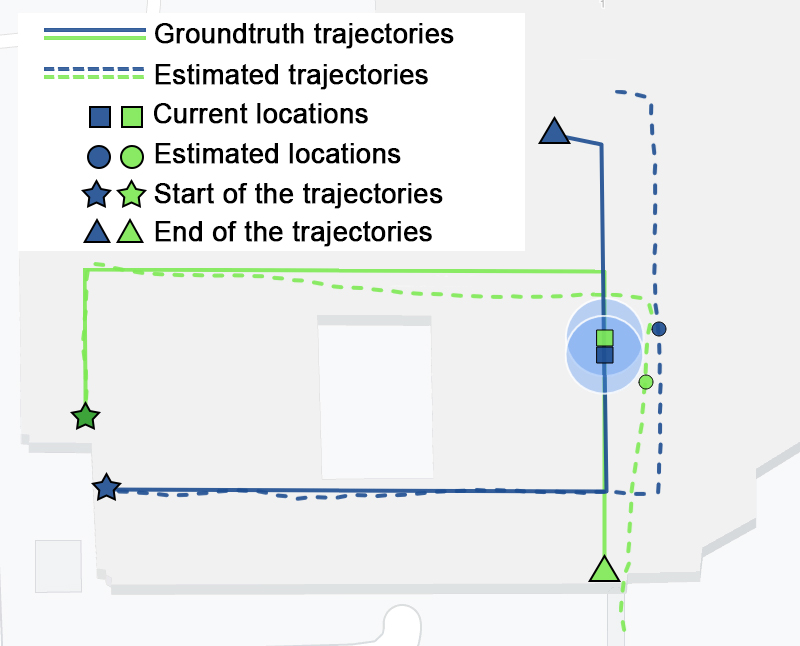}
    \caption{Devices within a collaboration range}
    \label{fig:collaborating-devices}
\end{figure}
Therefore, their main function, called \emph{handle\_matches}, receives as parameters a list of devices as objects, a timestamp representing the moment when the devices started collaborating, and two additional values representing the lower threshold of errors.
The lower threshold indicates the error above which a device considers its estimates divergent enough from the groundtruth and can, therefore, collaborate to improve them.
%
%
Figure~\ref{fig:collaborating-devices}, extracted from the orchestrator platform, illustrates two devices within a collaboration range with their groundtruth and estimated trajectories. We can see that the error on the estimated trajectories accumulates over time, thus leading to location estimates that diverge over time.

\subsection{Storage unit}

The storage unit stores raw measurements, trajectories, beacon maps, and parameters used for the experiments. 
It consists of disk storage and a relational database. The disk storage stores large files such as raw measurements, trajectories, and beacon maps.
The relational database offers a structured data organization to store and fetch information about the experiments. 
It connects the user via the orchestrator platform to the data stored on the disk. 
It also manages the users' privileges to the orchestrator platform, enabling the framework to be multi-user and multi-task. 

\subsection{Evaluation Process}

The evaluation process aims to compare the corrected trajectories with their corresponding estimated and groundtruth trajectories. 
Unlike other components listed in Figure~\ref{fig:architecture}, we consider the evaluation as a process that can be carried out independently of the framework. After the execution replay, the orchestrator platform provides data on trajectories, beacon locations and execution parameters in structured files (JSON, CSV) that researchers can use to evaluate their algorithms.
\section{Experiments and Execution Replay}
\label{sec:experiments}

This section describes the tracking algorithms built using \mobixim{}. It also shows how we collect raw measurements and replay participants' movements.

\subsection{Tracking algorithms}
We devise an inertial-based ITS incorporating an opportunistic error correction mechanism using nearby mobile devices' location estimates and locations of fixed BLE beacons. To achieve this, we start by filtering the inertial data to remove outliers associated with noise. Then, we run a state-of-the-art positioning algorithm to obtain the estimated trajectories. With the execution replay, we simulate peer-to-peer data exchanges and execute a collaborative algorithm. The tracking algorithms are described hereafter.

\myparagraph{Filtering algorithm.} 
As inertial sensors are prone to noise and generate outliers, we filter the data using a low-pass filter. This filter is designed to attenuate high-frequency signals above a cutoff frequency. 
We apply it to smooth the magnitude of acceleration computed using the following equation.
\begin{equation}
    mag = \sqrt{x^{2} + y^{2} + z^{2}}
\end{equation}
Where $mag$ is the magnitude of acceleration, $x$, $y$ and $z$ are the acceleration of a device on each of the corresponding three-axis.

We implement the algorithm in Python using the signal module of the SciPy library. Then, we integrate the code as a plugin into the framework.

\myparagraph{Positioning algorithm.}
After filtering the raw measurements, we use the PDR algorithm to estimate the participants' locations. 
First, we start by detecting steps in the magnitude of acceleration with a peak detection technique. We consider peaks in the acceleration amplitude as steps and trigger the PDR to detect the next location of a participant whenever a step is detected. The peak detection technique is also implemented using the SciPy library by analysing the acceleration magnitude as signals.

The PDR algorithm used at the core of the positioning algorithm is defined as follows.
\begin{equation}
    \begin{split}
    X_k = X_{k-1} + L * \cos({\theta})
    \\
    Y_k = Y_{k-1} + L * \sin({\theta})
    \end{split}
    \label{eq:pdr}
\end{equation}

Where the couple $(X_k, Y_k)$ represents the Cartesian coordinates of the next location computed using the previous coordinates $(X_{k-1}, Y_{k-1})$. $L$ is the step length of a user. We assume that users are aware of their step length, which they estimate by counting steps over a given distance. $\theta$ is the orientation obtained by the gyroscope.
The step length and the initial orientation of a device are parameters that can be adjusted on the orchestrator platform.
The resulting location is then converted to the WGS84 geographic coordinate system to construct an estimated trajectory.


\myparagraph{Collaborative algorithm.}
In our experiments, we use an enhanced version of a collaborative algorithm proposed in our previous research to correct the estimates of the PDR~\cite{diallo_decentralized_2024}.
The algorithm, defined in Algorithm~\ref{alg:aoe}, uses the inter-ranging distance between mobile devices as well as signals from nearby beacons to correct the estimated trajectories. 
When two mobile devices are close to each other in real life, they exchange their estimated errors and locations.
We estimate the error using an incremental value initially set to 0, which linearly increases over time. We decrease the error if the mobile device becomes stationary and starts collaborating with a peer or when it encounters a beacon, as depicted in Lines 10 and 16 of Algorithm~\ref{alg:aoe}.
We estimate that two devices are close to each other when they are below a distance of 4 meters. This distance respects social norms as it remains outside the area that individuals consider their personal space. Indeed, the \emph{theory of Proxemics} estimates the public space of individuals to be a circle of radius ranging between 3.6 and 7.6 meters~\cite{hall1968proxemics}.

When two mobile devices are within a collaboration distance, they draw a straight line between their two location estimates and position themselves at a distance on the straight line corresponding to the ratio of errors of each of the devices. Using a ratio of errors, we ensure that diverging devices, i.e. those with a large accumulated error, marginally affect devices with better estimates.

In addition to peer-to-peer collaboration, we also correct the estimates of mobile devices whenever they encounter fixed beacons. We estimate the distance between a mobile device and a beacon using the path loss model commonly used in the literature to model the relationship between RSSI and distance~\cite{debus_rf_2006}. The location of a mobile device is corrected to match the known location of a beacon only when the estimated distance between the mobile device and the beacon is below 2 meters. Our previous research shows that distance estimates are more reliable on short distances using BLE beacons~\cite{diallo_pragmatic_2024,diallo_decentralized_2024}. 

\begin{algorithm}[ht]
\caption{Drift correction with a mobile device or a beacon}
\label{alg:aoe}
\begin{algorithmic}[1]
    \State \textbf{Input:} devices: $A$ and $B$, lower-threshold $l$
    \State \textbf{Output:} $A$
    \If{$B.type$ is $``mobile"$ and $A.errors>l$}
        \State $sumErrors \gets A.errors + B.errors$
        \If{$sumErrors \neq 0$}
                \State $ratio \gets A.errors / sumErrors$
                \State $intermediatePoint \gets intermediatePoint(A.location, B.location, ratio)$
                    \State $A.location \gets intermediatePoint$
                \If{$A.location_{(t-1)} = A.location_{t}$}
                    \State $A.errors \gets A.errors-1$
                \EndIf
                
        \EndIf
    \EndIf
    \If{$B.type$ is $``beacon"$ and $l<A.errors$}
        \State $A.location \gets B.location$
        \State $A.errors \gets 0$
    \EndIf
    \State \textbf{return $A$}
\end{algorithmic}
\end{algorithm}

\subsection{Data collection}
To evaluate our proposed framework, participants collected raw measurements along representative groundtruths, i.e. based on the usual displacement of users. Our deployment environment is a single floorplan of a university building covering an area of up to 8400 m\textsuperscript{2} made of multiple rooms, corridors and facilities such as a restaurant, toilets and a library.
The data collection resulted in 45 trajectories spanning a cumulative distance of 5257 meters.

Participants collected data using six mobile devices: an iPhone 15 Pro, an iPhone 12 Pro, an iPhone 12, a Samsung Galaxy Tab S7, a Samsung Galaxy Tab A8, and a One-Plus Nord 2. These devices have different sensors, functionality, and prices.
For this experiment, we only need the inertial measurements to run the PDR algorithm. However, some mobile devices also collected signal measurements from physical beacons, creating a dense dataset that researchers could reuse to propose other tracking algorithms.
We position five virtual beacons, i.e. a density of 0.05 beacon per 100m\textsuperscript{2}, located at the most visited areas of the building. We use a hotspot detection algorithm to identify the ideal locations for positioning the beacons to maximize the chances of encountering mobile devices~\cite{diallo_pragmatic_2024}.

\subsection{Execution replay}
A major strength of \mobixim{} is the possibility for researchers to replay the movement of participants under multiple scenarios. For our experiments, we use the execution replay to combine all the collected trajectories into the same deployment environment before reproducing the movement of participants simultaneously to simulate a collaborative environment with data collected at different periods.
We also used the execution replay to adjust the number and location of beacons and observe how they affect the ITS's performance.
Finally, we executed movements associated with 45 trajectories collected using only six devices.
\section{Evaluation}
\label{sec:evaluation}

In this section, we measure the performance of the tracking algorithms. We introduce the evaluation metrics and then present our results before sharing the resources for reproducing the results and the experiments.

\subsection{Evaluation metrics}
\label{subsec:evaluation-metrics}

To evaluate the tracking algorithms, we compare each groundtruth trajectory with its corresponding estimated and corrected trajectories regarding similarity and positioning accuracy as described hereafter.

\myparagraph{Discrete Frechet Distance (DFD).}
The DFD, defined in Equation~\ref{eq:dfd}, is a measure of similarity used to compare two trajectories by considering the order and the location of each of their points.

\small
\begin{equation}
	\label{eq:dfd}
	dfd(i, j) = 
	\begin{cases}
		d(P_i, Q_j) & \text{if $i = j = 1$} \\
		max 
		\begin{cases}
			d(P_i, Q_j)  \\
			min 
			\begin{cases}
				dfd(i-1,j) \\
				dfd(i,j-1) \\
				dfd(i-1,j-1)
			\end{cases}
		\end{cases} & \text{otherwise}
	\end{cases}
\end{equation}
\normalsize

where $P$ and $Q$ represent trajectories such as $P = \langle p_1, ..., p_m \rangle$ and $Q = \langle q_1, ..., q_n \rangle$, with $p_i$ and $q_i$ representing points on each of the trajectories. $d(P_i, Q_j)$ is the ground distance $d$ between points pertaining to their respective trajectories $P$ and $Q$. The ground distance between two points $p_{l} = (\phi_{l}, \lambda_{l})$ and $q_{k} = (\phi_{k}, \lambda_{k})$ is computed with the Haversine formula defined in Equation~\ref{eq:haversine}.

\small
\begin{equation}
	\label{eq:haversine}
	d = 2 R \arcsin{\sqrt{\sin^{2}\left({\frac{\varphi_{l}-\varphi_{k}}{2}}\right)+\cos(\varphi_{k})\cos(\varphi_{l})\sin^{2}\left({\frac{\lambda_{l}-\lambda_{k}}{2}}\right)}}
\end{equation}
\normalsize
where $R$ is a constant representing the radius of Earth.

\myparagraph{Third Quartile of localization Errors.}
Also defined as \emph{positioning accuracy}, the third quartile of localization errors measures the 75th percentile of pairwise ground distances, defined in Equation~\ref{eq:haversine}, between each point in two given trajectories.
Compared to other metrics, such as the MSE or the MAE, the third quartile of localization errors is robust to outliers~\cite{potorti_comparing_2017}.

\subsection{Results}

\begin{figure}[ht]
    \centering
    \includegraphics[width=.85\linewidth]{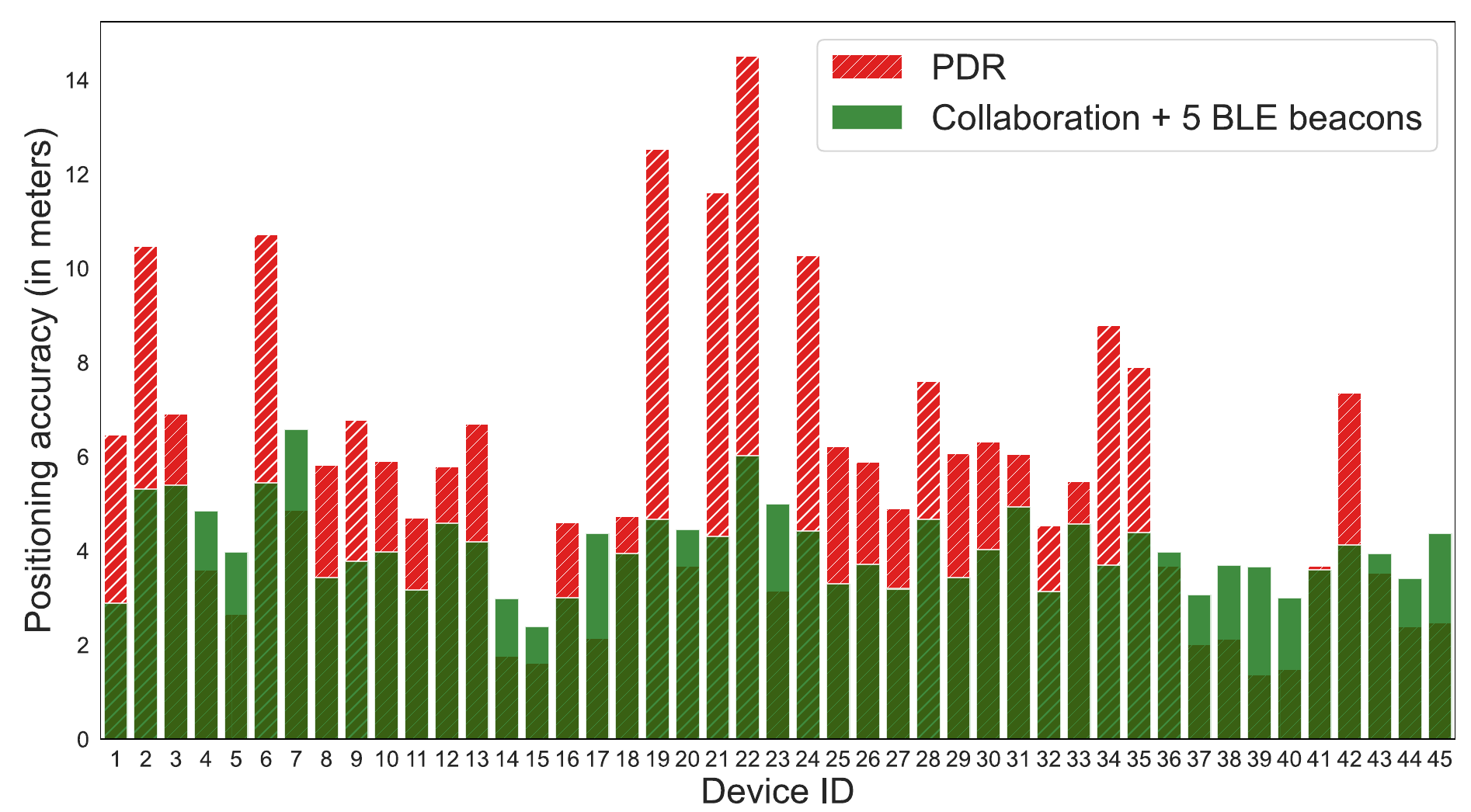}
    \caption{Third quartile of localization errors of all the devices}
    \label{fig:accuracy}
\end{figure}

As we observe in Figure~\ref{fig:accuracy}, our proposed algorithms significantly improve the positioning accuracy of 29 trajectories out of 45. In this figure, the red bars show the third quartile of localization errors of the positioning algorithm, and the green bars represent the results of the corrected algorithm. 
As expected, combining a collaborative algorithm with the PDR improves the estimates of trajectories that significantly deviate from their groundtruth. It is worth noting that some estimated trajectories with a high accuracy may experience minor degradation. However, in our experiments, the mean deviation observed was only 1.32 m for the affected trajectories. This degradation is negligible compared to the significant positive impact they have on the remaining trajectories.

\begin{figure}
    \centering
    \includegraphics[width=.9\linewidth]{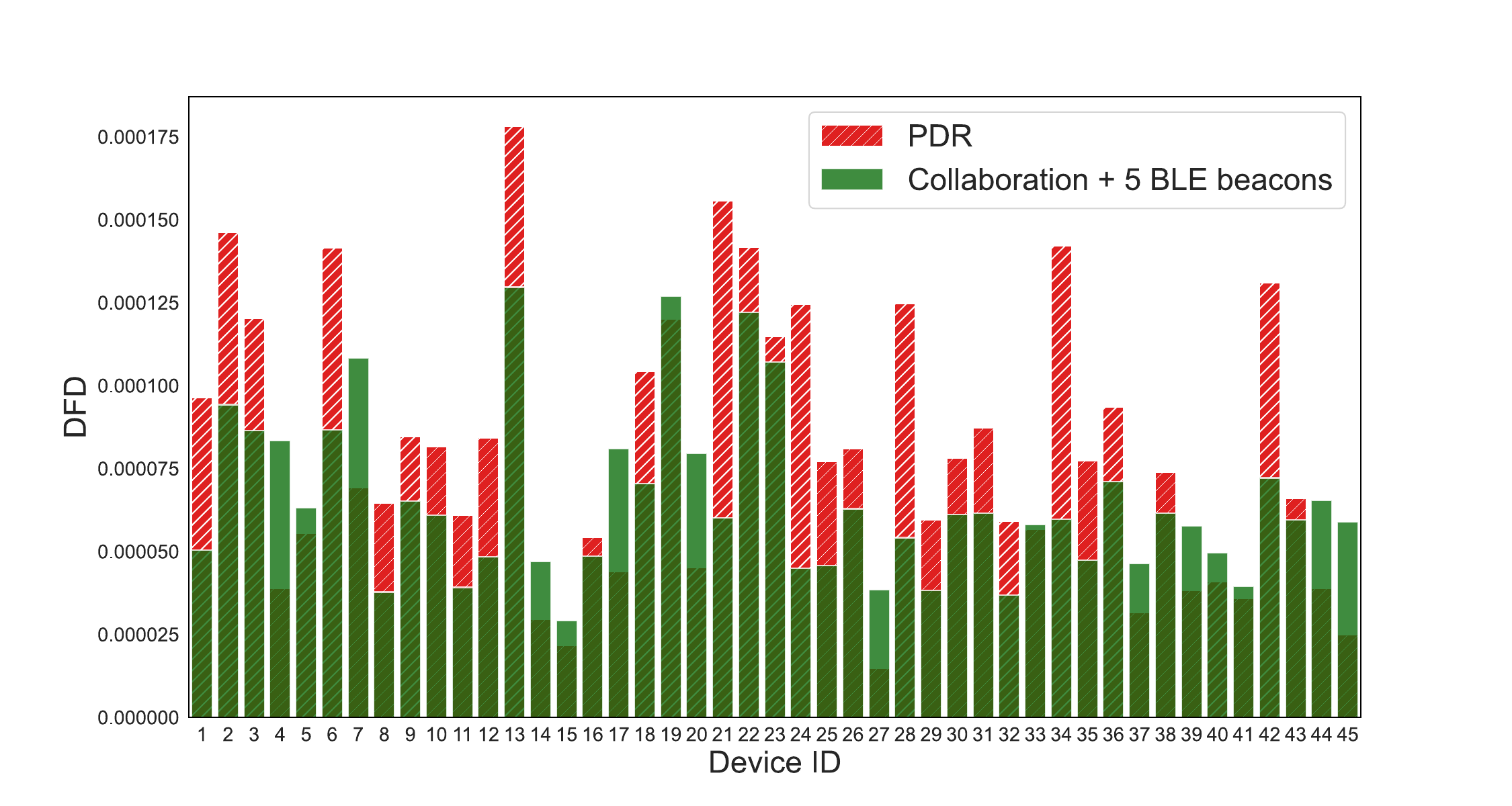}
    \caption{DFD score for all the devices}
    \label{fig:dfd}
\end{figure}

Our proposed algorithms also improve the similarity score of 29 trajectories out of 45. 
Figure~\ref{fig:dfd} shows the similarity score between the estimated and the corrected trajectories. A low score indicates a closer similarity with the groundtruth whereas a high score indicates divergence from the groundtruth.

Figure~\ref{fig:cdf} shows the Cumulative Distribution Function (CDF) of the localization errors for all the trajectories. The square dots are the third quartile of localization errors. 
With only five beacons coupled with peer-to-peer collaboration in a large deployment environment, we obtain an accuracy increase of up to 30\% with a mean positioning accuracy of 5.98 m for the estimated trajectories and 4 m for the corrected trajectories.

\subsection{Code and dataset}
A demo of the orchestrator platform is available on the following link: \url{https://doplab.unil.ch/mobixim}.
We preloaded the data used in the experiments to facilitate their reproducibility. Readers can access the dataset containing the raw measurements, the location of the beacons, the floorplan and the trajectories on the following link: \url{https://github.com/doplab/mobixim-evaluation}.

\section{Discussion and Future Work}
\label{sec:discussion}

In this section, we highlight the importance of incorporating collaborative ITS into \mobixim, we detail the impact of collaboration on the performance of the proposed tracking algorithms, and we conclude by presenting the limitations of the framework and the direction of our future research.

\subsection{Interest in collaborative ITS}

\begin{figure}
    \centering
    \includegraphics[width=0.46\linewidth]{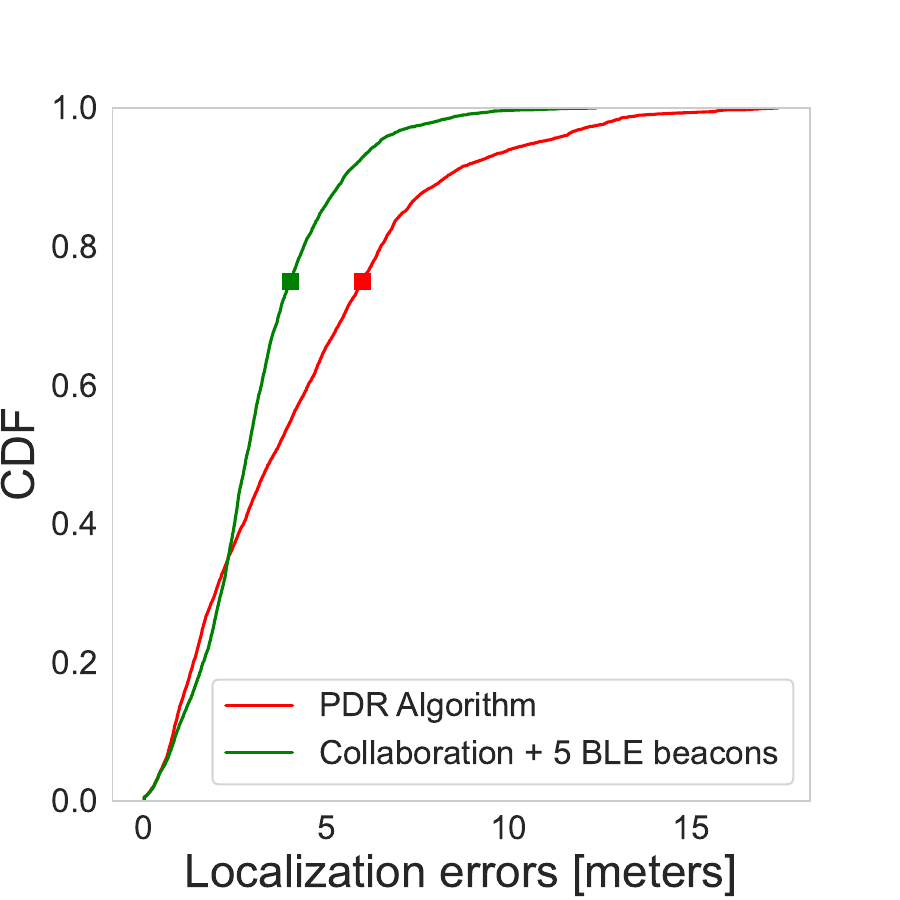}
    \caption{Cumulative Distribution Function (CDF) of the localization errors}
    \label{fig:cdf}
\end{figure}

One of the significant contributions of \mobixim over other frameworks is the integration of collaborative algorithms. 
Collaborative ITS have recently emerged after the massive interest in mobile contact tracing apps during the COVID-19 pandemic. 
%
%
By building an ITS that integrates a collaborative aspect, we emphasize the challenges of devising and evaluating collaborative tracking algorithms in a real-world environment.
Therefore, we introduce execution replay to reproduce the real movements of participants in a controlled environment to facilitate interactions with nearby mobile devices.
Additionally, \mobixim facilitates collaboration with nearby virtual and physical beacons to increase the accuracy of the ITS using RSSI. 
Integrating beacons is essential to enabling researchers to devise wireless-based ITS or to leverage the RSSI from beacons to further improve the accuracy of a collaborative ITS. 
Indeed, as presented in Figure~\ref{fig:cdf}, combining collaboration with a fixed beacon significantly improves the accuracy of a baseline positioning algorithm.

\begin{figure}[b] 
    \centering
    \includegraphics[width=1\linewidth]{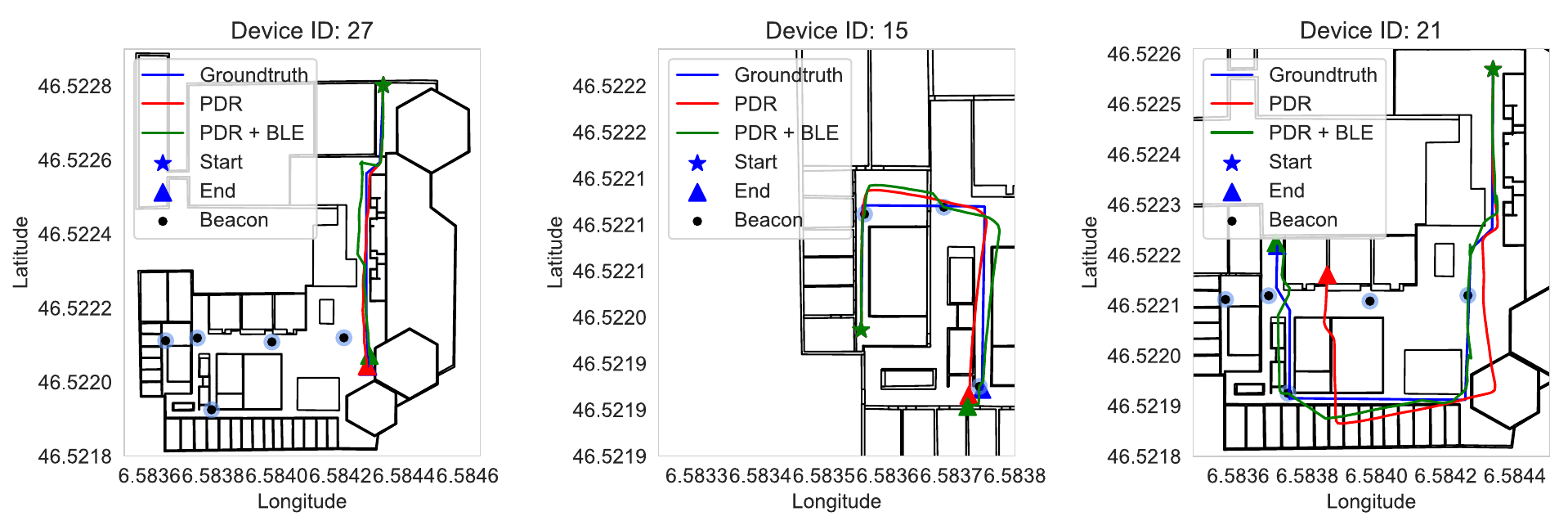}
    \caption{Sample of groundtruth, estimated and corrected trajectories in a single floor plans with virtual beacons}
    \label{fig:trajectories-sample}
\end{figure}

\subsection{Performance of the tracking algorithms}

Figure~\ref{fig:trajectories-sample} shows groundtruth, estimated and corrected trajectories of three devices. Device 27 made 91 collaborations with peers and never encountered any beacon. It improves its localization accuracy by 35\% compared to the PDR.
However, achieving multiple collaborations negatively impacts its similarity score, as it frequently uses nearby mobile devices' estimates to correct its location estimates.
Indeed, due to many collaborations, some devices frequently update their estimates, and their corrected trajectories follow irregular paths diverging from the groundtruth.

Device 15 collaborates only 36 times with peers and corrects its location estimates with three beacons. Due to the small number of collaborations, we observe a slight decrease in positioning accuracy and similarity score. 
For instance, collaborating with other devices slightly degrades its positioning accuracy from 1.62 m to 2.39 m, thus causing it to lose only 77 cm of positioning accuracy while improving peers' estimates.
Device 21 made 224 collaborations with peers and two corrections with beacons. By efficiently balancing the number of collaborations and corrections, it improved its similarity score by up to 61\% and its positioning accuracy by up to 63\%, from 11.6 m to 4.3 m.
Devices 15 and 21 correct their positions when they enter a beacon detection range. They position themselves at the beacon's position because we cannot reliably estimate the location of a mobile device solely based on the RSSI from a single beacon. 

\subsection{Limitations and future work}


To facilitate the data collection, we designed a mobile application that could be easily integrated into an ecosystem. We opted for a data collection process that requires participants to place markers on the ground to guide them along the groundtruth drawn on the orchestrator platform. However, we are aware that this approach can lead to errors in the order of a few centimetres or even a metre. This choice enables data to be obtained quickly to evaluate a tracking algorithm. For researchers looking for centimetre-level accuracy, Mobixim allows uploading their own data using the same format as the mobile companion app.

Some ITS are starting to incorporate techniques such as Angle-of-Arrival, which measures the slight phase differences of the Bluetooth signals across the anchor's multi-antenna array.
However, this is not yet ubiquitous because of a lack of hardware implementing this new Bluetooth 5.1 standard feature, as it requires specialized hardware embedding antenna array technology. In the future, we aim to integrate such hardware by extending the capabilities of the beacons as currently designed in \mobixim. 
\section{Conclusion}
\label{sec:conclusion}

We propose \mobixim, a framework for devising, evaluating and fine-tuning indoor tracking algorithms. We integrate a novel plugin architecture to ensure its extensibility, allowing researchers to reuse existing tracking algorithms. 
We evaluate it by designing an ITS that incorporates the three main types of ITS: inertial-based, wireless-based and collaborative. Using real-life inertial measurements, we substantially increase the accuracy of a baseline PDR algorithm. \mobixim{} helps collect data, set up a deployment environment, and replay movements, allowing researchers to visualize the real and estimated locations of participants and facilitate collaboration.
Additionally, we emphasize reproducibility by allowing researchers to exchange their data, tracking algorithms and parameters. 
In future versions of \mobixim, we aim to incorporate new indoor tracking algorithms based on the Angle-of-Arrival, as defined in recent specifications of the Bluetooth protocol.

\bibliographystyle{unsrt}  
\bibliography{references.bib}

\end{document}